\documentclass[12pt,draftclsnofoot,onecolumn]{IEEEtran}
\usepackage{blindtext}
\usepackage{amssymb}
\usepackage{amsmath}
\usepackage{graphicx}
\usepackage{cite}
\usepackage{citesort}
\usepackage{multicol}
\usepackage{amsfonts}
\usepackage{color}
\usepackage{subfigure}
\usepackage{epstopdf}
\usepackage{amssymb}
\usepackage{comment}
\usepackage{amsmath}
\allowdisplaybreaks[4]
\usepackage{booktabs}
\usepackage{multirow}
\usepackage{bmpsize}
\usepackage[ruled,linesnumbered]{algorithm2e}

\usepackage{lipsum}
\usepackage{stfloats}
\usepackage{lipsum}
\usepackage{stfloats}
\usepackage[ruled,linesnumbered]{algorithm2e}

\newtheorem{theorem}{\textbf{Theorem}}

\newtheorem{lemma}{\textbf{Lemma}}

\newtheorem{corollary}{\textbf{Corollary}}

\makeatletter
\def\ScaleIfNeeded{%
\ifdim\Gin@nat@width>\linewidth \linewidth \else \Gin@nat@width
\fi } \makeatother

\begin{document}
%\pagestyle{fancyplain}
%
%\pagestyle{fancy}
%\lhead[]%
 %   {\footnotesize Physical layer security}
%\cfoot

\title{\huge{Cell-free Terahertz Networks: A Spatial-spectral  Approach}}
\author{Zesheng Zhu, Lifeng Wang, Xin Wang, Bo Tan, and Shi Jin
%\thanks{Manuscript received May 22, 2022; revised August 22, 2022 and September 21, 2022; accepted September 28, 2022.  This work was supported
%in part by the National Key Research and Development Program under
%Grant 2021YFE0193300, and in part by Shanghai STCSM Program under
%Grant 22QA1401100. The work of Bo Tan was supported in part
%by the Academy of Finland under the project ACCESS (339519). The associate editor coordinating the review of
%this article and approving it for publication was M. Sepulcre. \emph{(Corresponding
%author: Lifeng Wang.)}}
\thanks{Z. Zhu, L. Wang, and X. Wang are with the School of Information Science and Engineering, Fudan University, Shanghai 200433, China (e-mail: $\rm\{lifengwang,xwang11\}@fudan.edu.cn$).}
\thanks{Bo Tan is with the Faculty of Information Technology and Communication Sciences, Tampere University, 33100 Tampere, Finland (e-mail:
bo.tan@tuni.fi).}
\thanks{Shi Jin is with the National Mobile Communications Research Laboratory,
Southeast University, Nanjing 210096, China (e-mail: jinshi@seu.edu.cn).}
}

\maketitle

\begin{abstract}
Cell-free network architecture plays a promising role in the terahertz (THz) networks since it provides better link reliability and uniformly good services for all the users compared to the co-located massive MIMO counterpart, and the spatial-spectral THz  link has the advantages of lower initial access latency and fast beam operations. To this end, this work studies cell-free spatial-spectral THz networks with leaky-wave antennas, to exploit the benefits of leveraging both cell-free and spatial-spectral THz technologies. By addressing the coupling effects between propagation angles and frequencies, we propose novel frequency-dependent THz transmit antenna selection schemes to maximize the transmission rate. Numerical results confirm that the proposed antenna selection schemes can  achieve much larger transmission rate than the maximal ratio transmission of using all the transmit antennas with equal subchannel bandwidth allocation in higher THz frequencies.
\end{abstract}
%\vspace{-0.20 cm}

%\vspace{-0.15 cm}
%========================================================================
\section{Introduction}
Immersive communications that support various extend reality (XR) applications are envisioned as one of 6G use cases~\cite{Sherman2023,huming2023}, which demand very large frequency bandwidths and low-latency transmissions. To enable those use cases, terahertz (THz) is the key radio technology to provide multi-GHz bandwidth above 100 GHz~\cite{ted_6G_2019}. However, the drawbacks of using conventional phased array for THz links are magnified in light of initial access latency~\cite{PoleseMag}, link budget~\cite{Rikkinen_THz_2020}, and beam management~\cite{Yuqiang_Heng2021,Li_Ruifu2022}. Some THz antenna solutions are studied including true-time delay (TTD) based phased array~\cite{A_J_Seeds1995}, phased subarray selection~\cite{Lin_chen2015} and leaky-wave antennas~\cite{D_headland2018,Knightly2023}, etc.

In millimeter wave networks with conventional phased arrays, beam management is time-consuming~\cite{PoleseMag,Yuqiang_Heng2021,Li_Ruifu2022}. Existing access procedure may lead to large initial access delay in THz links~\cite{PoleseMag}, which is further worsened by the high mobility of users~\cite{Yuqiang_Heng2021}. As such, spatial-spectral THz transmission seeks to reduce time consumption for beam management. The rationale behind it is that the propagation angles and frequencies are coupled and thus extensive beam training is unnecessary~\cite{Yasaman_2020,LZ2022}. TTD based phased arrays and leaky-wave antennas are two important alternatives for achieving spatial-spectral THz transmissions~\cite{Knightly2023}. While the former introduces variable delay lines to enable that classical phased arrays can also result in frequency-dependent beams~\cite{A_J_Seeds1995},  the latter is a cost-effective waveguide approach whose radiation efficiency is better than the phased array~\cite{D_headland2018}. In practice, TTD based phased arrays may have to overcome issues of making efficient THz phase shifters, feed network designs, and stringent requirement on the delay range of the circuit blocks~\cite{D_headland2018,Li_Ruifu2022}. As a low-cost and easy-to-manufacture traveling-wave antenna for making narrow beams, leaky-wave antenna has been applied in many areas such as frequency-division multiplexing THz communications~\cite{FDM_2015,LZ2022}, channel sounding~\cite{LZ_cl_2023}, sensing~\cite{Saeidi2021,LZ20222023}, and  physical layer security enhancement~\cite{Security2020,Chia-Yi_TFS}.

The coverage for an outdoor THz network involves not only antenna solution but also network architecture. {Traditional cellular architecture adopts co-located MIMO/massive MIMO antennas at base stations and users at the cell edge are subject to lower signal strengths, which may become much severer in THz frequencies due to blockages and molecular absorption losses~\cite{ted_6G_2019,LZ2022}.} Cell-free architecture deploys massive distributed single-antenna access points (APs), in which large numbers of APs are connected to a central processing unit~\cite{Ngo2017,emil2020CF}. Such a network architecture enables uniformly good services for all the users~\cite{Ngo2017,emil2020CF} and better link reliability~\cite{Arnold2021} compared to the cellular one. Moreover, cell-free THz links with shorter communication distances undergo lower atmospheric attenuation over distance (See Fig. 6 in~\cite{ted_6G_2019} and Fig. 5 in~\cite{LZ2022}). On the other hand, antenna solution in cell-free THz network needs to be delicately designed, since AP with single omnidirectional antenna is unable to support THz downlink transmission and directional beamforming is required to counteract the severe path loss. In fact, antenna solutions based on the channel state information of each transmit antenna element in conventional antenna arrays are not applicable in THz networks~\cite{Yasaman_2020}. Hence, existing works such as~\cite{Lin_chen2015} study the subarray-based channel estimation and signal processing schemes in the THz systems where each AP has multiple antenna subarrays, and each antenna subarray utilizes the analog beamforming to make narrow beam for large antenna gain.

Motivated by the advantages of  spatial-spectral transmission and cell-free network architecture in THz networks, we advocate the cell-free THz networks with leaky-wave antennas. {To the best of our knowledge, this is the first work to establish the cell-free THz networks with spatial-spectral coupling awareness. Novel frequency-dependent THz antenna selection schemes are designed to enhance the transmission rate and reduce the system overhead. In addition, optimal THz subchannel allocation algorithm is proposed to maximize the transmission rate, which has better performance than the equal allocation~\cite{Security2020} and co-located counterparts.}

\section{System Model}

In the large-scale cell-free THz downlink network with dense distributed APs, each AP is equipped with a TE$_1$ (lowest transverse-electric) mode leaky-wave antenna and user equipment (UE) has one omnidirectional receive antenna. Let $N$ denote the number of active APs that can serve the UE, and $x_{n}\left(f_i,\theta_n\right)$ ($n=1,\cdots,N$) is the binary indicator that the $n$-th AP is selected for downlink transmission at the $i$-th subchannel with the center frequency $f_i$; here, {$\theta_n$ is the line-of-sight (LoS) propagation angle (transmission direction).} The frequency $f_{\rm max}^n$ for obtaining maximum level of radiation by using a leaky-wave antenna at the $n$-th AP is~\cite{Yasaman_2020}
\begin{align}\label{sm_eq1}
f_{\rm max}^n=\frac{f_{\rm co}}{\sin(\theta_n)},
\end{align}
where $f_{\rm co}=\frac{c}{2\nu}$ is the cutoff frequency with the speed of light $c$ and leaky-wave antenna's inter-plate distance $\nu$. Therefore, the received signal strength (RSS) at the $i$-th subchannel by leveraging optimal coherent (matched filter) beamforming under equal power allocation in such a network is written as
{ \begin{align}\label{sm_SNR}
\hspace{-0.5 cm} {\small\widetilde{\gamma}\left(f_i\right)=\sum\limits_{n=1}^N {x_{n}\left(f_i,\theta_n\right) \gamma\left(f_i,\theta_n\right) \mathbf{1} \left(\gamma\left(f_i,\theta_n\right) \geq \gamma_{\rm th}\right)}},
\end{align}
where $ \gamma\left(f_i,\theta_n\right)$ is the RSS at the $i$-th subchannel for the selected AP $n$,  the indicator function $\mathbf{1} \left(\cdot\right)$ reflects that the THz link can be interrupted due to the lower signal strength conditions including higher pathloss and blockage effects}, {$\gamma_{\rm th}$ is the minimum RSS level (namely receiver sensitivity)}, and $\gamma\left(f_i,\theta_n\right)$ is given by
{\begin{align}\label{sm_SNR}
\gamma\left(f_i,\theta_n\right)=p \tilde G\left(f_i,\theta_n\right) \left| {{\hbar _{i,n}}} \right|^2,
\end{align}
where $p$ is each AP's transmit power spectral density (PSD),  $\tilde G\left(f_i,\theta_n\right) = \xi L{\rm{sinc}}\left[ {\left( { - \mathrm{j}\varrho  - {k_0}\left( f_i\right)\cos \theta_n  + \beta \left( f_i \right)} \right)\frac{L}{2}} \right]$ is the effective leaky-wave antenna gain with the radiation efficiency factor  $\xi$.} Here,  $L$ is the aperture length, $\mathrm{j}=\sqrt{-1}$, $\varrho$ is the attenuation coefficient, $k_0\left(f_i\right)=2\pi f_i/c$, $\beta\left(f_i\right) = k_0\left(f_i\right) \sqrt{1-\left(\frac{f_{\rm co}}{f_i}\right)^2}$, and $\left| {{\hbar _{i,n}}} \right|^2$ is the channel power gain of the $i$-th subchannel between the $n$-th AP and the UE. { As such, the frequency-dependent THz antenna selection with subchannel allocation is designed to maximize the transmission rate, accordingly, the problem is formulated as
\begin{align}\label{Opt_problem}
&{\mathop {\max }\limits_{{\bf{x,B,f}}} \sum\limits_{i} B_i \log_2\left(1+\frac{\widetilde{\gamma}\left(f_i\right)}{\sigma^2}\right)} \\
&\mathrm{s.t.} ~\mathrm{C1:}~\sum\limits_i {B_i }  \le B_{\rm total}, \nonumber\\
&\mathrm{C2:}~\bigcap\limits_{i} {\left\{ {f\left| {f \in \left[ {f_{i}  - \frac{{B_{i} }}{2},f_{i}  + \frac{{B_{i} }}{2}} \right]} \right.} \right\}}  = \emptyset , \nonumber\\
&\mathrm{C3:}~\left\| {\widetilde{\gamma } \left( {f_{i}  - \frac{{B_{i} }}{2}} \right)\left| {_{\rm dB} } \right. - \widetilde{\gamma } \left( {f_{i}  + \frac{{B_{i} }}{2}} \right)}\left| {_{\rm dB} } \right. \right\| < \varepsilon , \nonumber \\
&\mathrm{C4:}~x_{n}\left(f_i,\theta_n\right) \in \{ 0,1\},\; B_i \geq 0, \;~ f_{i} \geq 0, \;\,\;\forall i,\forall n, \nonumber
\end{align} }
where $\mathbf{x}=[x_{n}\left(f_i,\theta_n\right)]$, $\mathbf{f}=[f_i]$, $\mathbf{B}=[B_i]$ is the subchannel bandwidth vector, and $\sigma^2$ is the noise's PSD. Constraint $\mathrm{C1}$ is the frequency bandwidth limitation with the maximum value $B_{\rm total}$; $\mathrm{C2}$ ensures that there is no overlap between subchannels; {$\mathrm{C3}$ guarantees that  the RSS difference in the frequencies of a subchannel is less than a small value $\varepsilon$, which is also referred to as the coherence bandwidth of a subchannel.} { As seen in \eqref{Opt_problem}, each AP has the same subchannels with the same center frequencies.}

{The problem \eqref{Opt_problem} is combinatorial and highly non-linear, which is difficult to solve. As such,  we design the low-cost and efficient antenna selection schemes for determining $\{x_{n}\}$, and then propose a subchannel allocation algorithm for optimizing the center frequencies $\{f_{i}\}$ and  bandwidths $\{B_i\}$ of the subchannels in the following section.}

\section{Terahertz Transmit Antenna Selection with Subchannel Allocation}

From \eqref{Opt_problem}, it is intuitive that all the active APs need to be selected for maximizing the transmission rate, namely $x_{n}\left(f_i,\theta_n\right)=1, \forall i,\forall n$, which is referred to as maximal ratio transmission (MRT) scheme. However, the number of APs for maximal combining needs to be controlled in large-scale cell-free networks, in order to curtail the system overhead including the fronthaul/backhaul costs. In this section, four types of antenna selection schemes are investigated: 1) MRT; 2) Best $N_{\rm sel}$ transmit antennas; 3) Best transmit antenna; and 4) Nearest neighbor.
\subsection{Maximal Ratio Transmission}\label{MRT-1}
In the considered network, MRT can achieve the best performance but suffer the highest computational complexity. In this case, $\{x_{n}\left(f_i,\theta_n\right)=1\}$ is deterministic, and {problem \eqref{Opt_problem} reduces to a subchannel allocation problem, which is still combinatorial and non-convex. Moreover, subchannel bandwidth $B_i$ and its center frequency $f_{i}$ are coupled. To solve it,  we  develop a cross-entropy (CE) based algorithm to find optimal $\{f_{i}^*\}$ and $\{B_{i}^*\}$, which is detailed in \textbf{Algorithm 1}.}  CE method is an efficient approach to locate an optimal or near-optimal solution of combinatorial problems~\cite{BOTEV201}, which transforms the combinatorial problem as a rare-event estimation problem. In \textbf{Algorithm 1}, subchannels' center frequencies $\{f_{i}\}$ are deemed to be Gaussian distributions' parameters, and they are optimally evaluated by using maximum likelihood estimation (MLE). Specifically, the Gaussian distributions' parameters should be initialized to cover all the frequencies of interest $\left[f_{\rm co}, f^{\rm upper}\right]$, where $f^{\rm upper}$ is the upper bound of available THz frequency. {Given a center frequency sample, calculating its corresponding subchannel bandwidth is still hard, due to the non-overlap constraint $\mathrm{C2}$ and highly non-linear constraint $\mathrm{C3}$ in \eqref{Opt_problem}. Prior work~\cite{LZ2022} linearizes the leaky-wave antenna gain function $\tilde G\left(f_i,\theta_n\right)$ and approximately derives the subchannel bandwidths in the scenarios where the free-space path loss is dominant. As such, we compute the exact subchannel bandwidths $\{B_i\}$ by adopting the widely-used one-dimensional search to satisfy the constraints $\mathrm{C1}$--$\mathrm{C3}$ in \eqref{Opt_problem}. When there exists frequency overlap between adjacent subchannels, the overlapped frequency band is allocated to the subchannel with larger RSS at its center frequency.}
{  \begin{algorithm}[htp] \label{algorithmic1}
  Initialize Gaussian distributions' parameters {\small{$\left\{\mathcal{N}\left(\mu_i=f_{\rm co}+(i-\frac{1}{2})\frac{f^{\rm upper}-f_{\rm co}}{I}, \delta_i^2=16\frac{\left(f^{\rm upper}-f_{\rm co}\right)^2}{I^2}\right)\right\}_{i=1}^I$}} with the total number of subchannels $I$; the number of samples $m$, and the number of elite samples $m_{\rm elite}$  that have better objective values than other samples; the iteration index $t=0$\\
\While{$t < t^{\rm max}$}{i) Generate $\mathbf{S}_i \in \mathbb{R}^{m\times1}$ samples according to $\mathbf{S}_i\sim\mathcal{N}\left(\mu_i,\delta_i^2\right)$, then, given $\mathbf{S}_i$, obtain each sample's corresponding subchannel bandwidth  according to one-dimensional search under the constraints $\mathrm{C1}$--$\mathrm{C3}$;\\
ii) { Select $\mathbf{\widehat{S}}_i \in \mathbb{R}^{m_{\rm elite} \times1} $ elite samples from $\mathbf{S}_i$ that have better objective values of \eqref{Opt_problem} than other samples}, then, update the estimated parameters via MLE in a smoothing manner as follows:
{\small\begin{align*}
&\bar{\mu}_i=\frac{\sum\limits_{s=1}^{m_{\rm elite}} \mathbf{\widehat{S}}_i\left(s\right)}{m_{\rm elite}}, \bar{\delta}_i^2=\frac{\sum\limits_{s=1}^{m_{\rm elite}} \left(\mathbf{\widehat{S}}_i\left(s\right)-\bar{\mu}_i\right)^2}{m_{\rm elite}}, \\
&\mu_i(t+1)=\alpha \bar{\mu}_i+\left(1-\alpha\right) \mu_i(t),\\
&\delta_i^2(t+1)= \beta_t \bar{\delta}_i^2+\left(1-\beta_t\right)\delta_i^2(t),
\end{align*} }
where $\alpha$ and $\beta_t=\beta-\beta\left(1-1/(t+1)\right)^q$ are smoothing parameters;\\
iii) $t=t+1$;}
 {Optimal $\{f_{i}^*=\mu_i(t^{\rm max}),B_{i}^*(t^{\rm max})\}$ is obtained}
 \caption{\small{CE based Algorithm for Subchannel Allocation}}
\end{algorithm}
}
\subsection{Best $N_{\rm sel}$ Transmit Antennas}
The MRT antenna solution with all the active APs in subsection~\ref{MRT-1} creates high computational complexity and power consumption. To cut the system costs, we only select some of the best transmit antennas based on the RSS, i.e., $N_{\rm sel} \left(N_{\rm sel} < N\right)$ transmit antennas are selected according to the following rule:
\begin{align}\label{SBTA}
x_{n}\left(f_i,\theta_n\right)=1,\; \mathrm{if}\; \gamma\left(f_{\max}^n,\theta_n\right)=\gamma^{(n_{\rm sel})},
\end{align}
where $n_{\rm sel}=1,\cdots, N_{\rm sel}$, $f_{\max}^n$ is given by \eqref{sm_eq1} and $\gamma^{(n_{\rm sel})}$ is referred to as the $n_{\rm sel}$-th largest level of radiation among all the APs. Based on \eqref{SBTA}, we adopt the \textbf{Algorithm 1} with less numbers of APs to obtain the desired subchannel allocation.

\subsection{Best Transmit Antenna}
The best transmit antenna solution dictates that each subchannel only uses its best AP with the maximum RSS, to further cut the system overhead. As such, its antenna selection rule is
\begin{align}\label{BTA}
x_{n}\left(f_i,\theta_n\right)=1,\; \mathrm{if}\; \gamma\left(f_i,\theta_n\right)=\gamma_{\rm max}\left(f_i\right),
\end{align}
where $\gamma_{\rm max}\left(f_i\right)$ is the maximum RSS value at the $i$-th subchannel. In this case, the RSS of the $i$-th subchannel at the UE becomes
\begin{align}\label{sm_SNR_max_sel}
\hspace{-0.5 cm}\widetilde{\gamma}\left(f_i\right)=\mathop {\max }\limits_{n=1,\cdots,N} x_{n}\left(f_i,\theta_n\right) \gamma\left(f_i,\theta_n\right) \mathbf{1} \left(\gamma\left(f_i,\theta_n\right) \geq \gamma_{\rm th}\right).
\end{align}
Based on \eqref{BTA} and \eqref{sm_SNR_max_sel}, \textbf{Algorithm 1} is applied to attain the desired subchannel allocation.

\subsection{Nearest Neighbor}
Compared to the prior antenna selection solutions, the nearest neighbor  has the lowest system overhead since AP with the shortest communication distance is selected, namely
\begin{align}\label{NN1}
x_{n}\left(f_i,\theta_n\right)=1,\; \mathrm{if}\; d_n=\mathop {\min }\limits_{n=1,\cdots,N} d_{n},
\end{align}
where $d_n$ is the communication distance between AP $n$ and the UE. Based on \eqref{NN1}, the desired subchannel allocation is produced by using \textbf{Algorithm 1}. {It is noted that in the conventional cellular/cell-free networks without spatial-spectral coupling effects,  the nearest neighbor and the best transmit antenna solutions are almost identical since the antenna gain is independent of THz frequencies and the received signal strength is dominated by the pathloss.}

{The aforementioned four antenna selection schemes have the same computational complexity order when determining the optimal subchannels, since they all need to leverage \textbf{Algorithm 1} for solving the problem \eqref{Opt_problem}, and MRT scheme has no comparison operations since it employs all the APs. However, different antenna selection schemes lead to distinct fronthaul/backhaul costs, which depend on the scale of active APs in each antenna selection scheme. Therefore, MRT scheme has the highest fronthaul/backhaul costs.}

\section{Numerical Results}
This section presents numerical results to demonstrate the efficiency of the proposed antenna selection solutions. In the simulations, the basic system parameters are shown in Table~I; the communication distance is uniformly distributed within the coverage radius of 50m and the propagation angle from an AP to the UE is uniformly distributed, i.e, $\theta_n \in U(0, \frac{\pi}{2})$, and free-space path loss model is applied~\cite{LZ2022}; the leaky-wave antenna's attenuation coefficient is $\varrho=130$rad/m, aperture length is $L=0.09$m, and $\xi=1$; the smoothing parameters are set as $\alpha=0.8$, $\beta=0.7$, $q=5$, $I=40$, and the maximum number of iterations is $t^{\rm max}=30$ in \textbf{Algorithm 1}. Simulation results
are obtained by averaging over 2000 trials.

{\footnotesize{\begin{flushleft}
\begin{table}[h]\label{table12}
\centering
\caption{Simulation parameters}
\setlength\tabcolsep{3pt} \begin{tabular}{|l|c|}
  \hline
 Noise's PSD & ${\sigma}^2=-168$dBm/Hz\\ \hline
 Threshold for the minimum RSS & ${\gamma}_{\rm th}=-174.5$dBm/Hz\\ \hline
 RSS gap in the frequencies of a subchannel & $\varepsilon=0.5$dB \\ \hline
 Upper bound of available THz frequency & $f^{\rm upper}=300$GHz\\ \hline
Total transmit power  & $p_{\rm total}=p*B_{\rm total}=1$W \\
  \hline
\end{tabular}
\end{table}
\end{flushleft}}}

\begin{figure}
     \centering
    \subfigure[The cutoff frequency $f_{\rm co}=100$GHz.]{
         \centering
         \includegraphics[width=3.0 in,height=2.3 in]{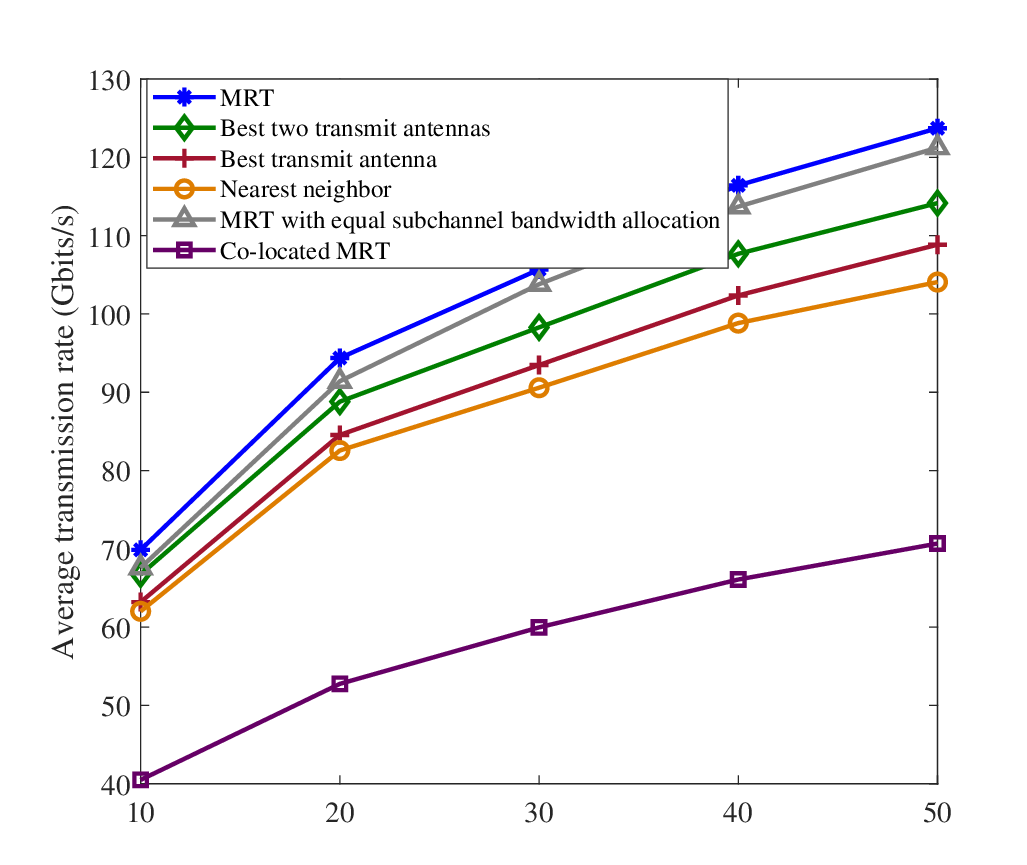}
      \label{figa1_k}
     }
     \subfigure[The cutoff frequency $f_{\rm co}=200$GHz.]{
         \centering
         \includegraphics[width=3.0 in,height=2.3 in]{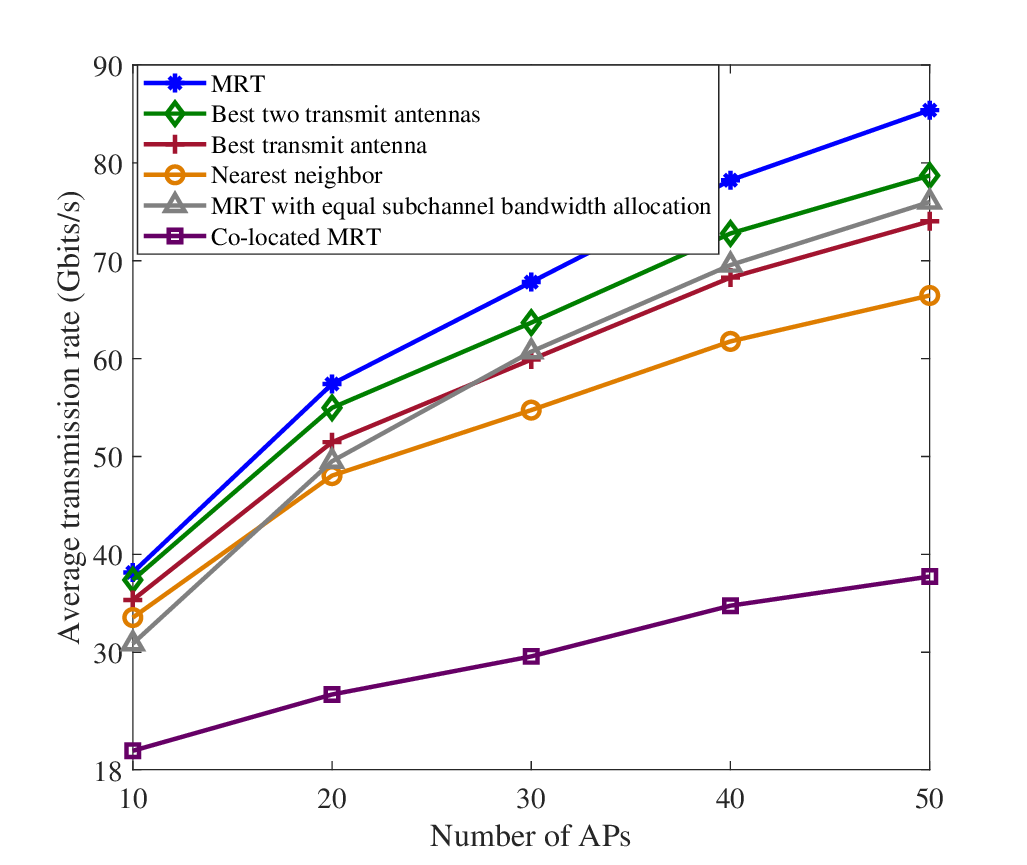}
      \label{figb1_k}
    }
 \caption{The average transmission rate versus number of APs under different cutoff frequency conditions with $B_{\rm total}=16$GHz and $N_{\rm sel}=2$.}
 \label{Fig_num_antenna}
\end{figure}

 Fig.~\ref{Fig_num_antenna} shows the average transmission rates achieved by the proposed antenna selections versus the number of APs under different cutoff frequency conditions. As mentioned before, the proposed MRT has the best performance, and  outperforms the baseline MRT with equal subchannel bandwidth allocation. { The best two transmit antennas scheme performs better than the best transmit antenna due to its larger antenna gain. Moreover, the performance of the low-complexity best transmit antenna scheme can be better than that of MRT with equal subchannel bandwidth allocation in higher THz frequencies as seen in Fig.~\ref{figb1_k}, since the RSS becomes very dynamic and significantly varies in higher THz frequencies, thus the subchannels need to be properly allocated, which further confirms the efficacy of the proposed subchannel allocation method in \textbf{Algorithm 1}.} The best transmit antenna scheme outperforms nearest neighbor since it has larger RSS.  Adding more APs in cell-free THz networks can significantly improve the transmission rate due to the large diversity gains, and using same bandwidth in lower THz frequencies (i.e., lower cutoff frequency is employed in Fig.~\ref{figa1_k} than Fig.~\ref{figb1_k}) has better performance due to the lower path loss.  The co-located MRT transmission achieves the lowest average transmission rate since it undergoes severe pathloss.

\begin{figure}
     \centering
    \subfigure[The cutoff frequency $f_{\rm co}=100$GHz.]{
         \centering
         \includegraphics[width=3.0 in,height=2.3 in]{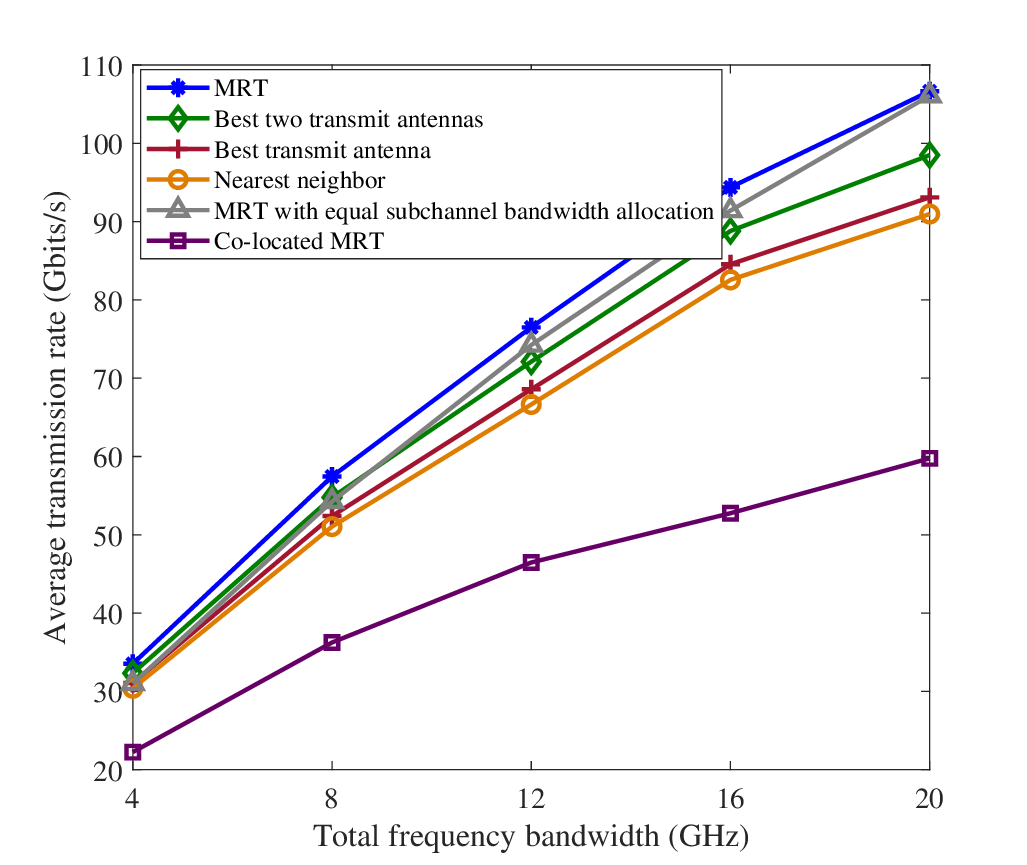}
      \label{figa2_k}
     }
     \subfigure[The cutoff frequency $f_{\rm co}=200$GHz.]{
         \centering
         \includegraphics[width=3.0 in,height=2.3 in]{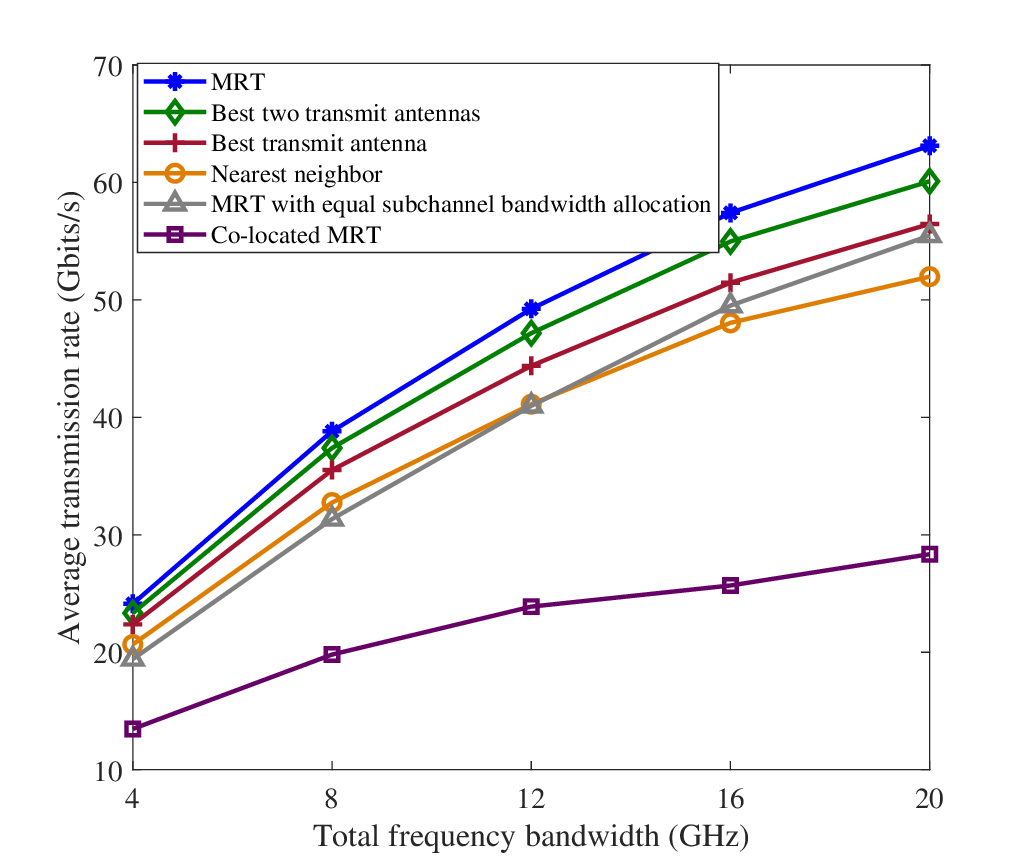}
      \label{figb2_k}
    }
 \caption{The average transmission rate versus total frequency bandwidth under different cutoff frequency conditions with $N=20$ and $N_{\rm sel}=2$.}
 \label{fig_num_BW}
\end{figure}

 Fig.~\ref{fig_num_BW} shows the average transmission rates achieved by the proposed antenna selections versus total frequency bandwidth  under different cutoff frequency conditions. As seen in Fig.~\ref{figa2_k}, in lower THz frequencies around $f_{\rm co}=100$GHz, the performance difference between the best transmit antenna and nearest neighbor is minimal since the path loss is dominant, and MRT of using all the APs with equal subchannel bandwidth allocation only performs better than the proposed best two (i.e., $N_{\rm sel}=2$) transmit antenna selection at very large frequency bandwidth values ($>16$ GHz in Fig.~\ref{figa2_k}). {In Fig.~\ref{figb2_k},   MRT of using all the APs with equal subchannel bandwidth allocation only performs better than the proposed nearest neighbor in higher THz frequencies around $f_{\rm co}=200$GHz with very large frequency bandwidth ($>16$ GHz in Fig.~\ref{figb2_k}),  due to the fact that the levels of RSS in higher THz frequencies fluctuate dramatically, in this case, subchannel allocation must be optimized by using the proposed \textbf{Algorithm 1}.} Again, we see that all the proposed antenna selection schemes in the cell-free THz networks perform much better than the co-located MRT.

\section{Conclusion}
Cell-free THz network with spatial-spectral coupling effects is a brand new network architecture with unique features, which could significantly reduce the complexity of beam management and system overhead. This letter proposed a cell-free THz network with leaky-wave antennas, in which four antenna selection schemes were developed to obtain efficient subchannel allocations with decreasing system costs, and their merits were confirmed by numerical results.

\bibliographystyle{IEEEtran}
\bibliography{mybib}

\begin{thebibliography}{}
\providecommand{\url}[1]{#1}
\csname url@samestyle\endcsname
\providecommand{\newblock}{\relax}
\providecommand{\bibinfo}[2]{#2}
\providecommand{\BIBentrySTDinterwordspacing}{\spaceskip=0pt\relax}
\providecommand{\BIBentryALTinterwordstretchfactor}{4}
\providecommand{\BIBentryALTinterwordspacing}{\spaceskip=\fontdimen2\font plus
\BIBentryALTinterwordstretchfactor\fontdimen3\font minus
  \fontdimen4\font\relax}
\providecommand{\BIBforeignlanguage}[2]{{%
\expandafter\ifx\csname l@#1\endcsname\relax
\typeout{** WARNING: IEEEtran.bst: No hyphenation pattern has been}%
\typeout{** loaded for the language `#1'. Using the pattern for}%
\typeout{** the default language instead.}%
\else
\language=\csname l@#1\endcsname
\fi
#2}}
\providecommand{\BIBdecl}{\relax}
\BIBdecl

\end{thebibliography}


\begin{thebibliography}{10}
\providecommand{\url}[1]{#1}
\csname url@samestyle\endcsname
\providecommand{\newblock}{\relax}
\providecommand{\bibinfo}[2]{#2}
\providecommand{\BIBentrySTDinterwordspacing}{\spaceskip=0pt\relax}
\providecommand{\BIBentryALTinterwordstretchfactor}{4}
\providecommand{\BIBentryALTinterwordspacing}{\spaceskip=\fontdimen2\font plus
\BIBentryALTinterwordstretchfactor\fontdimen3\font minus
  \fontdimen4\font\relax}
\providecommand{\BIBforeignlanguage}[2]{{%
\expandafter\ifx\csname l@#1\endcsname\relax
\typeout{** WARNING: IEEEtran.bst: No hyphenation pattern has been}%
\typeout{** loaded for the language `#1'. Using the pattern for}%
\typeout{** the default language instead.}%
\else
\language=\csname l@#1\endcsname
\fi
#2}}
\providecommand{\BIBdecl}{\relax}
\BIBdecl

\bibitem{Sherman2023}
X. Shen, J. Gao, M. Li, C. Zhou, S. Hu, M. He, and W. Zhuang, ``Toward
  immersive communications in {6G},'' \emph{Front. Comput. Sci.}, vol. 4, 2023.

\bibitem{huming2023}
M.~Hu, L.~Wang, B.~Tan, and S.~Jin, ``Two-tier 360-degree video delivery
  control in multiuser immersive communications systems,'' \emph{{IEEE} Trans.
  Veh. Technol.}, vol.~72, no.~3, pp. 4119--4123, Mar. 2023.

\bibitem{ted_6G_2019}
T.~S. Rappaport \emph{et~al.}, ``Wireless communications and applications above
  100 {GHz}: {O}pportunities and challenges for {6G} and beyond,'' \emph{IEEE
  Access}, vol.~7, pp. 78\,729--78\,757, June 2019.

\bibitem{PoleseMag}
M. Polese, J. M. Jornet, T. Melodia, and M. Zorzi, ``Toward end-to-end,
  full-stack 6G terahertz networks,'' \emph{{IEEE} Commun. Mag.}, vol.~58,
  no.~11, pp. 48--54, Nov. 2020.

\bibitem{Rikkinen_THz_2020}
K.~Rikkinen, P.~Ky\"{o}ti, M.~E. Leinonen, M.~Berg, and A.~P\"{a}rssinen,
  ``{THz} radio communication: {L}ink budget analysis toward {6G},''
  \emph{{IEEE} Commun. Mag.}, vol.~58, no.~11, pp. 22--27, Nov. 2020.

\bibitem{Yuqiang_Heng2021}
Y.~Heng, J.~G. Andrews, J.~Mo, V.~Va, A.~Ali, B.~L. Ng, and J.~C. Zhang, ``Six
  key challenges for beam management in {5.5G} and {6G} systems,'' \emph{{IEEE}
  Commun. Mag.}, vol.~59, no.~7, pp. 74--79, July 2021.

\bibitem{Li_Ruifu2022}
R.~Li, H.~Yan, and D.~Cabric, ``Rainbow-link: {B}eam-alignment-free and
  grant-free {mmW} multiple access using true-time-delay array,'' \emph{{IEEE}
  J. Sel. Areas Commun.}, vol.~40, no.~5, pp. 1692--1705, May 2022.

\bibitem{A_J_Seeds1995}
I. Frigyes and A. J. Seeds, ``Optically generated true-time delay in
  phased-array antennas,'' \emph{ IEEE Trans. Microw. Theory Techn.}, vol. 43,
  no. 9, pp. 2378--2386, Sept. 1995.

\bibitem{Lin_chen2015}
C.~Lin and G.~Y. Li, ``Adaptive beamforming with resource allocation for
  distance-aware multi-user indoor terahertz communications,'' \emph{{IEEE}
  Trans. Commun.}, vol.~63, no.~8, pp. 2985--2995, Aug. 2015.

\bibitem{D_headland2018}
D. Headland, Y. Monnai, D. Abbott, C. Fumeaux, and W. Withayachumnankul,
  ``Tutorial: Terahertz beamforming, from concepts to realizations,'' \emph{APL
  Photonics}, vol. 3, no. 5, pp. 1--32, 2018.

\bibitem{Knightly2023}
J. M. Jornet, E. W. Knightly, and D. M. Mittleman, ``Wireless communications
  sensing and security above 100 {GHz},'' \emph{Nat. Commun.}, vol. 14, 841,
  pp. 1-10, Feb. 2023.

\bibitem{Yasaman_2020}
Y.~Ghasempour, C.~Yeh, R.~Shrestha, D.~Mittleman, and E.~Knightly, ``Single
  shot single antenna path discovery in {THz} networks,'' in \emph{Proc.
  MobiCom}, 2020, pp. 1--13.

\bibitem{LZ2022}
Z.~Lin, L.~Wang, B.~Tan, and X.~Li, ``Spatial-spectral terahertz networks,''
  \emph{{IEEE} Trans. Wireless Commun.}, vol.~21, no.~6, pp. 3881--3892, June
  2022.

\bibitem{FDM_2015}
N.~J. Karl, R.~W. McKinney, Y.~Monnai, R.~Mendis, and D.~M. Mittleman,
  ``Frequency-division multiplexing in the terahertz range using a leaky-wave
  antenna,'' \emph{Nature Photonics}, vol.~9, pp. 717--721, Sept. 2015.

\bibitem{LZ_cl_2023}
Z.~Lin, L.~Wang, J.~Ding, B.~Tan, and S.~Jin, ``Channel power gain estimation
  for terahertz vehicle-to-infrastructure networks,'' \emph{IEEE Commun.
  Lett.}, vol.~27, no.~1, pp. 155--159, Jan. 2023.

\bibitem{Saeidi2021}
H.~Saeidi, S.~Venkatesh, X.~Lu, and K.~Sengupta, ``{THz} prism: {O}ne-shot
  simultaneous localization of multiple wireless nodes with leaky-wave {THz}
  antennas and transceivers in {CMOS},'' \emph{IEEE J. Solid-State Circuits},
  vol.~56, no.~12, pp. 3840--3854, Dec. 2021.

\bibitem{LZ20222023}
Z.~Lin, L.~Wang, J.~Ding, Y.~Xu, and B.~Tan, ``Tracking and transmission design
  in terahertz {V2I} networks,'' \emph{{IEEE} Trans. Wireless Commun.},
  vol.~22, no.~6, pp. 3586--3598, June 2023.

\bibitem{Security2020}
C.-Y. Yeh, Y.~Ghasempour, Y.~Amarasinghe, D.~M. Mittleman, and E.~W. Knightly,
  ``Security in terahertz {WLANs} with leaky wave antennas,'' in \emph{Proc.
  ACM WiSec}, 2020, pp. 1--11.

\bibitem{Chia-Yi_TFS}
C.-Y. Yeh, A.~Cohen, R.~G.~L. D'~Oliveira, M.~M\'{e}dard, D.~M. Mittleman, and
  E.~W. Knightly, ``Securing angularly dispersive terahertz links with
  coding,'' \emph{{IEEE} Trans. Inf. Forensics Security}, vol.~18, pp.
  3546--3560, Apr. 2023.

\bibitem{Ngo2017}
H.~Q. Ngo, A.~Ashikhmin, H.~Yang, E.~G. Larsson, and T.~L. Marzetta,
  ``Cell-free massive {MIMO} versus small cells,'' \emph{{IEEE} Trans. Wireless
  Commun.}, vol.~16, no.~3, pp. 1834--1850, 2017.

\bibitem{emil2020CF}
E.~Bj\"{o}rnson and L.~Sanguinetti, ``Making cell-free massive {MIMO}
  competitive with {MMSE} processing and centralized implementation,''
  \emph{{IEEE} Trans. Wireless Commun.}, vol.~19, no.~1, pp. 77--90, Jan. 2020.

\bibitem{Arnold2021}
M.~Arnold, P.~Baracca, T.~Wild, F.~Schaich, and S.~t. Brink, ``Measured
  distributed vs co-located massive {MIMO} in industry 4.0 environments,'' in
  \emph{Proc. EuCNC/6G Summit}, 2021, pp. 306--310.

\bibitem{BOTEV201}
Z. I. Botev, D. P. Kroese, R. Y. Rubinstein, and P. L. Ecuyer, ``Chapter 3-the
  cross-entropy method for optimization,'' \emph{Handbook of Statistics}.
  Elsevier, 2013, vol. 31, pp. 35-59.

\end{thebibliography}

\end{document}